%% file: main.tex
\documentclass[letterpaper,twocolumn,10pt]{article}
\usepackage{hyperref}
\usepackage{usenix,epsfig,endnotes}
\usepackage{svg}
\usepackage{amsmath}
\usepackage{lipsum}

\usepackage{caption}
\usepackage{subcaption}

\usepackage{listings}
\begin{document}

\date{}

\title{Asynchronous I/O --- With Great Power Comes Great Responsibility}

\author{
{\rm Constantin Pestka}\\
German Aerospace Center
\and
{\rm Marcus Paradies}\\
LMU Munich
\and
{\rm Matthias Pohl}\\
German Aerospace Center
}

\newcommand{\IOENGINE}{\textsc{Ringbearer}}
\newcommand{\IOURING}{\texttt{io\_uring}}
\newcommand{\WINIORING}{\texttt{I/O Rings}}
\newcommand{\SPDK}{\texttt{SPDK}}
\newcommand{\XNVME}{\texttt{xNVMe}}
\newcommand{\NVME}{\texttt{NVMe}}
\newcommand{\DPDK}{\texttt{DPDK}}
\newcommand{\LIBAIO}{\texttt{libaio}}
\newcommand{\ZIG}{\texttt{Zig}}
\newcommand{\PCIE}{\texttt{PCIe}}
\newcommand{\AIO}{\texttt{aio}}

\newcommand{\sref}[1]{§~\ref{#1}}

\newcommand{\SUBMITCMD}{\texttt{Submit()}}
\newcommand{\POLLCMD}{\texttt{PollRequest()}}
\newcommand{\BLOCKCMD}{\texttt{BlockOnRequest()}}

\maketitle


\input{inc/content/abstract} 
\input{inc/content/introduction}
\input{inc/content/background}
\input{inc/content/challenges}
\input{inc/content/architectures}
\input{inc/content/summary}

{\footnotesize \bibliographystyle{acm}
\bibliography{inc/bib/bib}}

\end{document}

%% file: inc/content/abstract.tex
\begin{abstract}
The performance of storage hardware has improved vastly recently, leaving the traditional I/O stack incapable of exploiting these gains due to increasingly large relative overheads.
Newer asynchronous I/O APIs, such as \IOURING{}, have significantly improved performance by reducing such overheads, but exhibit limited adoption in practice.
In this paper, we discuss the complexities that the usage of these contemporary I/O APIs introduces to applications, which we believe are mostly responsible for their low adoption rate.
Finally, we share implications and trade offs made by architectures that may be used to integrate asynchronous I/O into DB applications.
\end{abstract}

%% file: inc/content/introduction.tex
\section{Introduction}
\label{sec:Intro}

SSDs have witnessed substantial performance advancements in recent years.
Even commodity SSDs can achieve low double-digit $\mu$s latency, double-digit GiB/s throughput, and can process millions of IOPS~\cite{kioxiaCd8p, phisonGen5Controller,samsungPm1743}. 
These performance advances pose tremendous challenges to contemporary I/O software stacks.
Overheads, such as frequent context switches in blocking I/O, have made it increasingly difficult to fully utilize modern hardware efficiently.
Consequently, there has been a variety of novel I/O APIs that attempt to address these performance issues, i.e., \NVME{}~\cite{nvmeSpecs} at the protocol level, 
\IOURING{}~\cite{iouringIntro}
and \WINIORING{}~\cite{winIoRings} at the OS level, \SPDK{}~\cite{Yang17} for OS bypassing access, and \XNVME{}~\cite{xNvme} as a unifying abstraction layer on top.
One unifying design aspect of these APIs is the transition from a blocking to an asynchronous, completion-based interface.
Depending on the used I/O API and configuration, this communication mechanism largely replaces the need for syscalls with mostly lock-less inter-thread communication between multiple kernel and user-space threads.
Thus, these new async I/O APIs have the potential to be significantly more efficient than their blocking counterparts~\cite{Leis18,iouringIntro,IOApICmp,IoAPICmp2,Haas23}.
However, shifting to these APIs involves moving from a blocking, preemptive multitasking design to an explicitly async, cooperative multitasking approach.
This change transfers the responsibility for managing the time between I/O request submission and receiving the results to the application.
This may not pose an issue for applications that are already multi-threaded, e.g., by using an async job system or an explicit event loop.
However, it may require significant refactoring for simpler multi-threaded or single-threaded applications.

While it is possible to issue low MIOPS per thread (cf.~Figure~\ref{fig:IOPSvsSMT},~\cite{iouringIntro}), the sheer extent to which the throughput capability of I/O devices has grown in recent years, vastly outpaces the capability of a single thread on essentially all contemporary platforms.
Current server generations are frequently equipped with up to 128 \PCIE{} 5.0 lanes that can accommodate dozens of SSDs, each capable of serving multiple MIOPS~\cite{amdEpycGen4}.
Achieving a saturation of 10-20\% on such a system fundamentally requires multiple dedicated threads worth of CPU time.
Applications that wish to fully utilize modern I/O hardware thus have to implement some degree of parallelism on top of these APIs.
Unfortunately, implementing parallelism is generally non-trivial (§~\ref{sec:banes_of_async},~\ref{sec:arch}), 
especially while retaining acceptable levels of performance.
Despite the significant performance potential the adoption of modern async I/O APIs in I/O heavy applications is low.
Many state-of-the-art I/O libraries, such as libuv, seastar and tokio~\cite{libuv,seastar,tokioURing} have only limited experimental support and widely used libraries, such as libc and 
libc++,
do not support novel I/O APIs such as \IOURING{} at all.
The few libraries that do support these, integrate them in a relatively limited, experimental fashion~\cite{Zig, tokioURing}. 
Similarly, while async I/O APIs have been explored in academic DBMSs~\cite{Haas23, asyncIoCoroutines} and commercial DBMSs~\cite{tigerbeetle}, the majority of large, commercial DBMSs still rely on blocking I/O APIs.
We believe this is due to the many, often complex responsibilities that these APIs bring to applications (e.g., the requirement to implement efficient parallelism and user-space task scheduling).

In this paper we will first in~\sref{sec:why_move_to_async} briefly present an overview of the benefits of async I/O and will then in~\sref{sec:banes_of_async} in detail go over the not so frequently discussed challenges users of async I/O have to cope with.
In~\sref{sec:arch}, we conclude with a discussion on a set of architectural patterns that can be applied for the purpose of using async I/O in applications and how they relate to the previously discussed challenges.
In this discussion, we will include classifications of existing systems that already utilize async I/O and discuss trade-offs made in their implementations.

%% file: inc/content/background.tex
\section{Background}
\label{sec:why_move_to_async}

The design of blocking I/O APIs dates back to a time when storage was slow and mostly based on magnetic media.
Multiple decades of CPU and storage hardware evolution have continuously eroded the initial assumptions made during the design of traditional blocking I/O APIs.
On the CPU side, single-core performance improvements have stagnated, as its main historical driver of frequency scaling has slowed notably.
Current CPU performance has mostly improved due to IPC gains and is largely limited by the memory subsystem, which itself heavily depends on its caching hierarchy.
This is particularly critical for context switches, as caches have to be flushed to avoid side channels and thus additionally cause high indirect costs \cite{syscallsBadMicrokernel, syscallsBadSyscallAlternative, syscallsBadMicrokernel2,Zhou23}.
On the storage side, modern SSDs, while experiencing similar frequency scaling issues and thus exhibiting lower gains in latency, provide massive I/O throughput by exploiting a high degree of internal parallelism.
In the following we provide an overview of the challenges imposed by blocking I/O on modern hardware and how async APIs attempt to address these.


\subsection{The Inefficiencies of Blocking I/O}
\label{sec:blocking_is_bad}

\begin{figure}[t!]
\centering
\includegraphics[width=1.0\columnwidth]{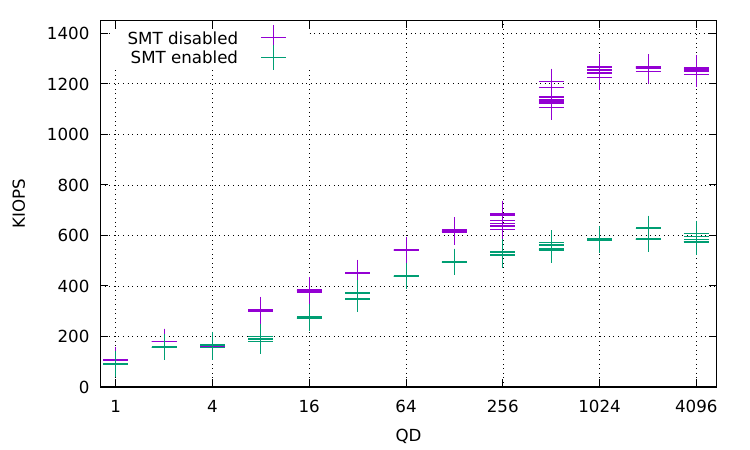}
\vspace{-20pt}
\caption{IOPS queue depth scaling. Uses \IOURING{}, direct I/O, SQ and I/O poll, 1M 4K sequential reads per run, 10 runs each, preceded by preconditioning run. Run on a Ryzen 3900X, Samsung 990 Pro 1TB and Linux 6.8.0-39.}
\label{fig:IOPSvsSMT}
\vspace{-14pt}
\end{figure}

Blocking I/O APIs delegate the responsibility to schedule work between I/O request submission \& completion to the OS rather than the application.
To access the OS scheduler used for this purpose and to access the kernel I/O stack, user space threads have to invoke a syscall to submit any I/O operation.
The resulting, at the very least two, context switches are a the fundamental performance issue of blocking I/O APIs.
Another issue is that the process, which issues a blocking I/O request, can only make further forward progress if it posses an additional, currently not running, not already blocked thread, to have a \emph{chance} to be scheduled to run upon submitting the blocking I/O request.
Further, while depending on the configuration, the process is often more likely to do so the more runable threads its poses.
This can lead to over-subscription, which can be disadvantageous for overall system performance and efficiency.
While heavily depending on configuration knobs, e.g., thread priorities, scheduling classes \& cgroup configuration, the mostly stochastic nature of the OS scheduling can also be undesirable as it might lead to unpredictable behavior.
Over-subscription can be problematic in the context of modern SSDs, as exploiting their vast bandwidth is predicated on a sufficiently high number of concurrently in-flight requests (cf.~Figure~\ref{fig:IOPSvsSMT}).
To achieve this an application can either utilize additional threads or use vectored I/O.
Vectored I/O allows the submission of multiple I/O requests via a single syscall.
While this reduces the required syscalls per I/O operation, it does not allow for different I/O request types to be mixed and requires all requests to be known at the time of submission.
The latter can be problematic in applications where the need for I/O requests does not occur in large batches or is unpredictable.
Further, while in the multi-threaded approach, the application can always immediately utilize the result of every I/O request once it completes.
In the vectored approach, the thread generally remains blocked until all results arrive.


\subsection{The Boons of Async I/O}
\label{sec:boons_of_async}

Most I/O operations are fundamentally asynchronous.
Blocking I/O APIs try to hide this fundamentally async nature and the accompanying complexity from the user.
Instead, async I/O APIs fully expose this to the user.
This splits the, for the user, essentially atomic I/O operation of blocking APIs into two distinct events, the submission and completion, and moves the scheduling of tasks for the interim into user space, allowing most syscalls be eliminated.
While some old I/O APIs (e.g., \AIO{}) have not taken this opportunity and consequently have been criticized~\cite{iouringIntro}, newer I/O APIs (e.g., \IOURING{}, \WINIORING{}, or \SPDK{}) have done so.
They typically achieve this via two lockless ring buffers, the \emph{Submission Queue} (SQ) through which I/O requests are submitted, and the \emph{Completion Queue} through which notifications of completion are received.
These two communication channels can vastly reduce or completely alleviate the need for syscalls and replace them with lower overhead atomic memory operations.
The exact details of when syscalls are still required highly depend on configuration parameters, such as using \emph{submission queue} and \emph{I/O device polling} (SQ and IO poll), but have significant impact on performance~\cite{iouringIntro, IOApICmp, IoAPICmp2}.
Another advantage of async APIs is that applications can utilize the higher degree of available context to make more optimal scheduling decisions.
Instead of relying on the coarse per thread prioritization schemes of the OS scheduler, applications may optimize their scheduling for generally higher throughput, latency or QoS or choose to prioritize specific groups of tasks, such as persisting critical logs.
For SSDs the most critical benefit is the improved efficiency of achieving a high number of concurrent IOPS.
Here, async APIs combine the advantages of the threaded, blocking approach and vectored I/O, while not suffering from their limitations.
Results of individual requests can always be utilized upon arrival and are not limited by batching.
Further amortizing costs via batching is possible, but failing to do so is significantly less costly as the overall submission overhead is vastly reduced.


%% file: inc/content/challenges.tex
\section{The Banes of Async I/O}
\label{sec:banes_of_async}



While, async I/O is conceptually rather simple and provides many opportunities to improve the efficiency of I/O-intensive applications, actually doing so is anything but trivial for all but the simplest applications.
Utilizing these APIs, especially when scaling parallelism to fully utilize modern hardware, introduces complex challenges that we believe have not been sufficiently addressed so far.

\subsection{I/O has become CPU bound}
\label{sec:Io_CPU_bound}

\begin{figure}[t!]
\centering
\includegraphics[width=1.0\columnwidth]{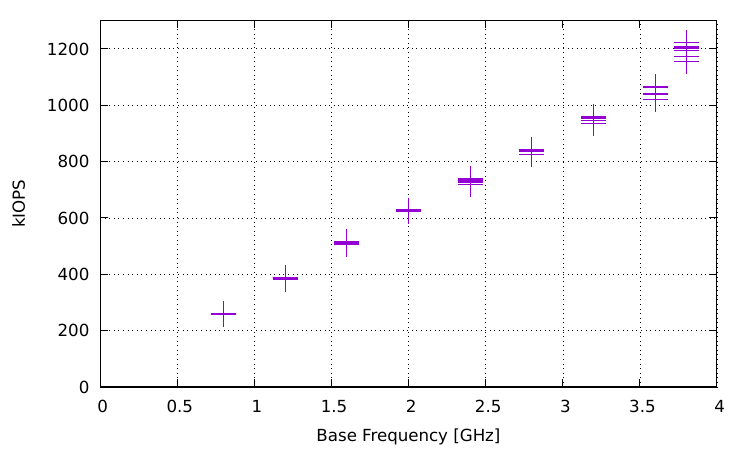}
\vspace{-20pt}
\caption{Frequency set in the BIOS (Parameters cf.~Figure~\ref{fig:IOPSvsSMT})}
\label{fig:IOPSvsFreq}
\vspace{-15pt}
\end{figure}


While historically I/O performance has been limited by I/O device speeds, recently it has become increasingly CPU-limited, and lead to the development of novel async I/O APIs.
And while these APIs have massively improved CPU efficiency and hence also maximally achievable performance on one thread~\cite{iouringIntro,IoAPICmp2,Haas23,Leis18,IOApICmp}, even with these APIs, I/O performance on one thread is still closely linked to CPU performance (cf.~Figure~\ref{fig:IOPSvsFreq})
The CPU time of a single thread is not sufficient to saturate modest modern hardware setups, such as a single \PCIE{} 5.0 SSD, even in idealized benchmark scenarios.
The first implication of the degree to which I/O performance is CPU-limited on modern hardware is that a significant degree of parallelism via multi-threading is strictly required to saturate contemporary hardware setups.
The second implication is that the budget of CPU time both for handling I/O related tasks and any other tasks, is limited.
Even short delays, e.g., executing other application logic, processing the result of finished IOPs or a few LLC misses, can reduce performance by orders of magnitude (cf.~Figure~\ref{fig:IOPSvoCallbackLat}).
Techniques to retain high performance on modern CPUs \& memory subsystems, include utilizing small, non-fragmented, cache-efficient objects \& data structures and custom memory allocators.
Although they are in heavy use in some communities, such as the Linux kernel, they are not yet commonly used in I/O heavy applications.
Additionally, while SSD performance characteristics themselves are non-trivial due to internal factors (e.g., page size, garbage collection, SLC caching), on top of these, the factors that impact general CPU performance now also impact I/O performance.
This includes software induced factors, e.g. cache misses, but also setup related factors, such as dynamic frequency scaling, which is strongly dependent on thermal headroom and the power management settings.
Among these, we found that SMT has a pronounced impact on performance (cf.~Figure~\ref{fig:IOPSvsSMT}).
On modern hardware these factors have to be considered both for I/O heavy applications and in I/O related benchmarks, not only due to their impact on performance, but also due to the associated understandability and reproducibility issues.


\subsection{Moving Scheduling to User Space}
\label{sec:sched_to_userspace}

While we have previously touched on the advantages of moving scheduling responsibilities to user space, this also has notable downsides.
Scheduling itself is a famously non-trivial topic.
This, especially in combination with the performance requirements and necessity of multi-threading, introduces an additional burden on the application implementations.
This is only complicated further, as the application has to be able to efficiently cope with the async nature of the APIs, e.g., via async job systems that run on top of a thread pool (\sref{sec:exec_arch}).
Furthermore, the discontinuous nature of the logical control flow of async tasks can complicate debugging, as often a full callstack does not exist.


\subsection{Per API Instance Overheads}
\label{sec:Api_overhead}

\begin{figure}[t!]
  \centering
  \includegraphics[width=1.0\columnwidth]{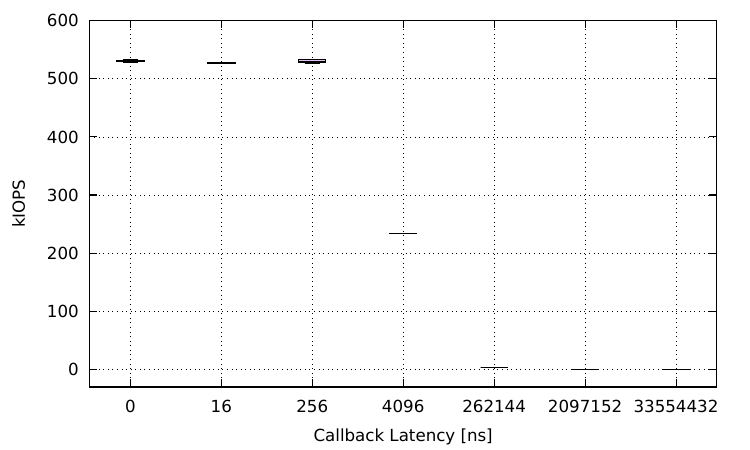}
  \vspace{-20pt}
  \caption{IOPS scaling with post I/O operation callback latency (Parameters cf.~Figure~\ref{fig:IOPSvsSMT}, but using random reads)}
  \label{fig:IOPSvoCallbackLat}
  \vspace{-15pt}
\end{figure}

One of the significant differences between OS-level and OS-bypassing APIs is that instances of the former can come with a steep upkeep cost, most notably due to SQ polling kernel threads.
Similar to I/O polling, which removes the need for expensive IRQs to handle arriving results at the cost of additional CPU resources, SQ polling removes the need for syscalls during submission at the cost of a dedicated thread per API instance.
While optional, both polling modes significantly increase performance~\cite{iouringIntro, IOApICmp, IoAPICmp2} at the cost of the additional CPU time.
As these threads are implicitly controlled via a $ms$ granularity timeout, they remain active even if only few requests are submitted to them in small enough time intervals.
This can be problematic in architectures, where it is difficult or impossible to match the number of API Instances to the hardware capabilities.
Failing to adapt the active number of API instances to a varying I/O load can, depending on the workload, be similarly detrimental.
The significant upkeep cost of the polling threads generally imposes the challenge for architectures to balance the overheads introduced by sharing API instances with the cost of low API instance utilization.


\subsection{Task Division and Task Distribution}
\label{sec:task_div_dist}



Two key architectural decisions are how logical tasks are divided into subtasks (\sref{sec:task_partition}) and how the execution of these subtasks is organized among threads (\sref{sec:exec_arch}).
There is a trade-off between schemes that divide at a smaller granularity increasing dispatch overheads, such as function calls, persisting metadata of requests while they are in flight, and potentially inter-thread communication and schemes dividing at a larger granularity, which makes utilizing dedicated I/O threads (\sref{sec:io_thread_or_not}) impossible.
Similarly, the scheme chosen to distribute requests among threads has to balance the amount of inter-thread communication and the utilization of API instances.

\subsection{To I/O Thread or not to I/O Thread}
\label{sec:io_thread_or_not}

It is a consequential question whether or not all I/O-related tasks, such as the submission \& polling for completions, managing the metadata of in-flight requests, and handling partial completions, should be isolated into separated subtasks that are then executed on dedicated \emph{I/O threads}.
Not doing so has the advantage that it avoids the need for additional inter-thread communication.
However, executing any non-I/O related subtask on the same thread that submits I/O (\emph{inline execution}) will reduce the API instance utilization to a degree proportional to the execution time of the non-I/O tasks executed.
Given the extent to which throughput and hence API instance utilization drops based on the execution time of tasks executed inline (cf.~Figure~\ref{fig:IOPSvoCallbackLat}), CPU efficiency will drop significantly if inline execution is used alongside I/O or SQ polling, even for short non-I/O tasks of a few $\mu s$.

\subsection{Increased complexity over traditional I/O}
\label{sec:io_uring_complexity}

Async APIs (e.g., \IOURING{}) are more complex than blocking I/O, as they include more features and introduce additional responsibilities to the user.
In the following we discuss some of them.
As the SQ and CQ of \IOURING{}'s API instances are \emph{statically} sized ring buffers, the application has to avoid overflowing the SQ.
Enforcing ordering and atomicity among I/O operations is more involved, especially if cancellations are involved, which also makes the semantics of fsync more complex than usual.
\IOURING{} exposes features, such as multi-shot accepts, registered file descriptors \& buffers, which can improve efficiency, but also increase complexity.
Finally, to correctly use an API instance with more than one thread, familiarity with modern memory models is required to achieve correctness.
Furthermore, most programmers use high-level libraries, such as 
libc
to perform I/O.
However, the vast majority of I/O libraries do not support the new async I/O APIs and the few that do (e.g.,~\cite{tokioURing, Zig}) do so in a limited fashion.
While an exhaustive discussion of these libraries is out of the scope of this paper, limitations such as utilizing a single-threaded event loop and, thus, limited throughput limit their scope of usability on modern hardware.
The complexities of a correct and efficient implementation that utilizes e.g. \IOURING{} is thus highly relevant, as most applications that wish to utilize these more efficient I/O APIs will have to implement the integration on their own.

%% file: inc/content/architectures.tex
\section{Async I/O Architectures}
\label{sec:arch}

\subsection{Task Partitioning Schemes}
\label{sec:task_partition}



There are three common ways to partition a task including async I/O.
To compare these, it is helpful to divide a task into subtasks separated by I/O operations.
We then define a \emph{tasklet} as a sequence of instructions including one or more sub-tasks that one scheduled run from start to finish without interruption on the same thread.

\subsubsection{Full Partitioning}
\label{sec:full_partition}

In \emph{Full Partitioning}, every subtask is divided into its own tasklet.
If there is a subsequent IOP, then the submission of the according request is appended to the tasklet of the previous subtask.
The polling for completion of an IOP is encapsulated into a separate tasklet, which spawns itself again on failure and the tasklet of its successor subtask on success.
This approach is highly flexible, as it allows implementations to choose any distribution among threads and prioritization scheme.

\subsubsection{Callback Partitioning}
\label{sec:callback_part}



Another method similar to full partitioning is to use a callback function that is called after the polling of an I/O request returns a successful status.
This method combines the tasklet for polling I/O requests with the subsequent subtask.
It should be noted that this implies that said sub-task is also executed by the same thread that performed the polling of the I/O request, hence preventing the usage of I/O threads, which is not the case for full partitioning.

\subsubsection{Coroutines}
\label{sec:coroutine_part}

Coroutines are functions that can be suspended and resumed at predefined points, here most notably after IO submission and unsuccessful polling for completion and thus can be used to encapsulate an entire task.
They are typically implemented as stackless coroutines, which consist of a heap-allocated stackframe containing the function arguments, current locals, and a state enum.
When a coroutine is called the state enum is used to determine the point at which execution is to be resumed.
Some implementations enforce a specific execution architecture, such as a single-threaded eventloop, thus also always introducing the limitations of said architecture, such as limited parallelism.
The properties of coroutines thus vary wildly between implementations of different languages such as C++, Rust, JavaScript, Zig, Go or manual implementations~\cite{manualAsync} and are hence difficult to discuss in this context.
One notable performance implication of stackless coroutines is that resuming coroutines, especially nested ones, can be rather expensive, as coroutines frames are heap allocated and often large and thus hostile to caches, as they must fit the entire current state of the full task.
As polling for I/O completions must remain inexpensive, using coroutines for async I/O can be problematic.
A notable upside of coroutines is that they encapsulate more context of the current task in the callstack as they contain the full task, which can be beneficial for debugging.
Coroutines thus have been a popular choice for async I/O \cite{asyncIoCoroutines} or other async operations like hiding pre-fetching operations~\cite{corobase, manualAsync}.

\subsection{Execution Architectures}
\label{sec:exec_arch}

\begin{figure}[t!]
  \includegraphics[width=1.0\columnwidth]{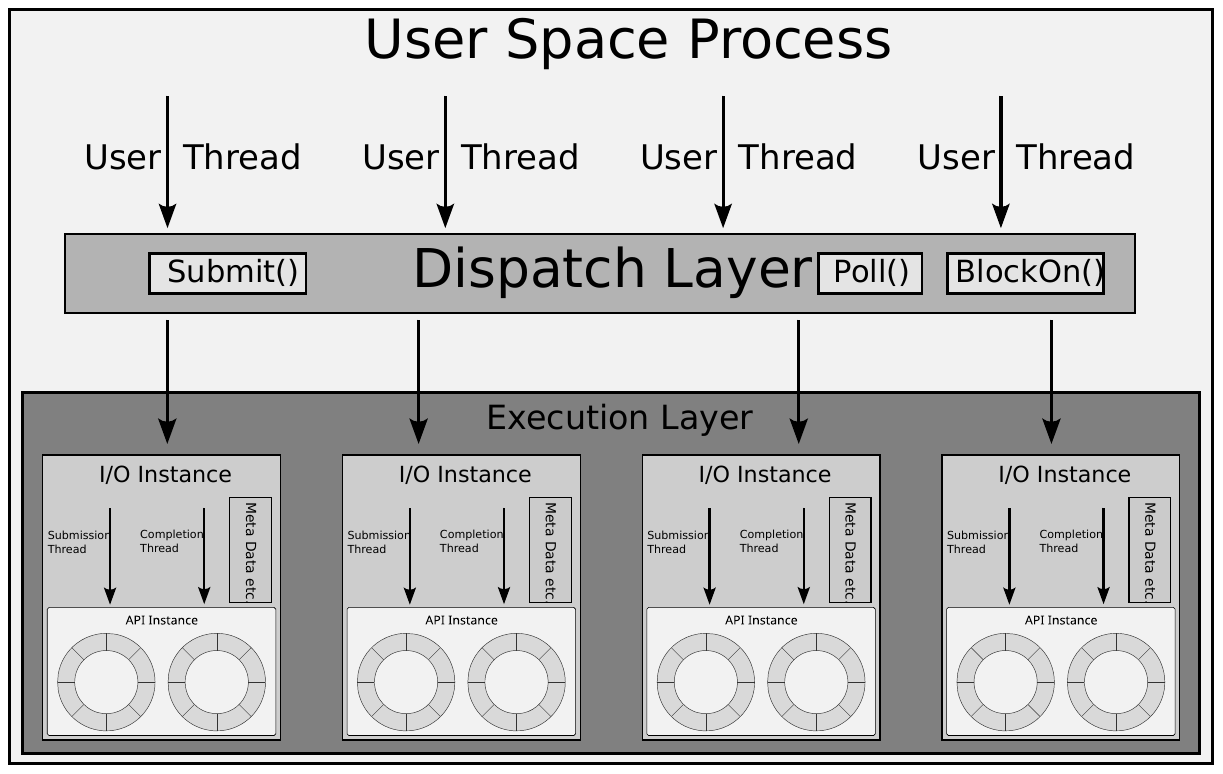}
  \vspace{-20pt}
  \caption{Static I/O thread pool}
  \vspace{-15pt}
  \label{fig:pool}
\end{figure}

In this chapter, we discuss the distribution of tasklets among threads for execution.
Our discussion of architectures will focus on their ability to scale efficiently to higher degrees of parallelism.
Due to this we will exclude fully single threaded applications and single threaded event loops, that can be encountered in e.g. the Zig standard library or the JS runtime, from our discussion, as their limited throughput is insufficient for modern hardware.
Similarly, we will limit the discussion to \emph{thread per core} architectures and thread per core compatible architectures, as the alternative approach of over-subscription and preemptive multi-tasking are problematic in this context due to their often excessive amount of context switches, whose problematic nature has been discussed previously.







\subsubsection{Direct Access or M-to-N sharing}
\label{sec:arch_m_n_sharing}


The significant upkeep cost of API instances is a strong incentive to utilize them efficiently.
A simple strategy to achieve this is to share them between multiple threads, e.g. by allowing all of the user space threads ($N$) of an application to access any of the API instances ($M$) directly.
This has the advantage that the resources allocated for I/O and non-I/O tasks can be adjusted by changing $M$ and $N$, respectively.
While this mostly decouples the amount of resources allocated for I/O from the the ones allocated for other purposes, as the CPU time to perform e.g. submissions is still provided by the $N$ user space threads, the chosen configuration is static and cannot be adapted later.
In scenarios with a low I/O load, the submission of requests can thus not be dynamically restricted to a smaller set of API instances resulting in low CPU efficiency (\sref{sec:Api_overhead}).
Additionally, this architecture does not provide the capability to buffer requests if a thread attempts to submit to an API instance with a full SQ and thus must provide this itself.
Additionally, it is the responsibility of each thread to choose an API instance to submit to, which could cause congestion issues.
The most relevant property of this architecture for our purposes is, however, its issues with scalability.
Not only are more expensive synchronization primitives required
compared to architectures that restrict access to a specific API instance to only one dedicated thread, but more importantly, the potential for contention is significantly higher by design than for the other presented architectures.
This is especially relevant for the polling operations on the CQs.
Likely due to said performance issues we are not aware of any project utilizing this architecture.




\subsubsection{Shared Nothing}
\label{sec:arch_thread_per_core}


In this architecture, each user-space thread submits any of its I/O requests to its private API instance(s).
This architecture proposes to achieve parallelism via multiple fully independent instances of the single-threaded architecture.
This approach has a few significant upsides, in particular its simplicity and requiring \emph{no} inter-thread communication.

However, this architecture is not applicable in its pure form in most real-world applications.
While it works remarkably well in benchmarks, where requests are generated right before submission, and results are discarded immediately upon completion, this effectively ignores that in most real applications, there is logic to be executed that causes an I/O operation and processes its result.
In most applications chains of logic operations, often triggered by non-compile-time known events such as read input from a file or received network packet, eventually require one or more I/O OPs to be issued, of which the output is then processed further.
This architecture assumes that the workload \emph{can} be efficiently partitioned into tasks that are then executed fully independently from each other.
In other words, this assumes the task is at least approximately embarrassingly parallel and can be efficiently statically scheduled.
The effectively non-deterministic latency of I/O, especially for network I/O, conflicts with this assumption.
Most I/O intensive applications, like DBMSs or web servers are inherently dynamic, with numerous compile-time unknowns and often complex dependencies.
The resulting complexity and interconnectedness of their control flow graphs are fundamentally at odds with being easily and efficiently partitionable.
That not all or even most tasks are embarrassingly parallel is a well-known fact in many domains, such as GPGPUs.

Another property of this architecture is that, due to not communicating among threads, I/O throughput available to a task is limited to the maximum throughput of one thread.
Further implications are due to utilizing inline execution (\sref{sec:io_thread_or_not}), especially regarding CPU efficiency.
This is especially critical in this architecture, as there is a hard coupling between available resources for non-I/O and I/O tasks.
This is due to the fact that every user space thread must have at least one dedicated API instance and thus, if SQ polling is used, comes with a dedicated polling thread.
This severely restricts the maximum amount available resources for non-I/O tasks, while not resorting to over-subscription, as e.g. 32 user space threads, that are also still required to perform the I/O related work, submission, polling etc., would necessitate 32 kernel threads even when using the minimum of one API instance per thread.

\subsubsection{Static I/O Thread pool}
\label{sec:arch_static_pool}


This architecture (cf.~Figure~\ref{fig:pool}) attempts to address both, the inter-thread syn\-chro\-ni\-zation-based performance issues of the direct access architecture and the restrictions with respect to the applicability of the shared-nothing architecture.  
To achieve this, it adds a dispatch layer to which all user-space threads submit requests and receive an object from that can be polled for the status of the submitted request.
The distribution layer distributes the submitted requests to one of potentially multiple \emph{I/O instances}.
An I/O instance consists of either two threads, one responsible for submission one for completion, or one thread responsible for both, as well as one or more I/O API instances that are exclusive to said thread(s).
This design moves the area in which inter-thread communication is required from the API instances to the dispatch layer.
An advantage over the direct access architecture is the massively reduced degree of contention on the API instances, as only ever a single thread will access the SQ or CQ, respectively, which is especially crucial for the CQ, due to the polling nature of the access.
Concurrent polling for completions of requests from multiple users thus does not reduce I/O throughput, as users poll on the objects they received for this purpose and not directly on the CQ.
Furthermore, while significant contention could occur during submission to the dispatch layer, the design of the dispatch layer can be adapted to address this if required, e.g. by increasing the number of data structures used to buffer and distribute the requests.

While resources allocated for I/O and non-I/O can be scaled independently, similar to the direct access architecture, but decoupled further as the CPU time for I/O tasks is provided by the dedicated threads of the I/O instances, this allocation is still static.
Furthermore, as the threads of I/O Instance are dedicated threads, it is possible to utilize these as I/O threads.
Implementations could also as an additional optional mechanism choose to provide the option to utilize inline execution via callbacks to prioritize latency-sensitive tasks.
While this architecture provides many benefits the potentially significant increase in inter-thread communication compared to the shared-nothing approach remains its most significant trade-off.
The extent of this overhead depends on the granularity of submitted tasks and on the many implementation details of the dispatch layer and I/O instances.
These design decisions range from the used data structures and synchronization primitives in the dispatch layer over the used scheduling scheme to the mechanism used to manage the metadata of in flight requests.

A system that can be categorized in this architecture is Go's scheduler for \emph{goroutines}, which are, like KSEs, stackful, thus reducing the performance issues due to nesting.
Go's usability in the context of high performance async I/O is, however, debatable, due to its managed nature, most notably, but not limited to its garbage collector.
An academic system that similarly should be classified under this architecture is Leanstore~\cite{Haas23}, which utilizes a single queue as the dispatch layer and callback task partitioning.
Merzljak et al.~\cite{asyncIoCoroutines} base their work on Leanstore, thus sharing its architecture, but utilize C++ coroutines instead.

\subsubsection{Dynamic I/O Threadpool}
\label{sec:dyn_pool}

Notably, Haas et al. discussed the impact of the SQ poll threads in their design and concluded that it would be more beneficial to turn off the feature entirely~\cite{Haas23}.
We argue that given the extent to which both SQ polling can increase performance~\cite{iouringIntro, IOApICmp, IoAPICmp2} further exploration of the management of these kernel threads is warranted.
If the design is adapted such that the amount of I/O resources, including these kernel threads, are accurately scaled to what is actually required, both statically to the hardware capabilities and dynamically to the current I/O load, the performance gains of SQ polling could be leveraged while retaining good or even better CPU efficiency.
To this end two issues of the previous architecture would need to be addressed.
The first is to utilize full task partitioning as the primary partitioning scheme over callback partitioning or coroutines.
Besides the established trade-offs, this allows the amount of resources allocated to I/O to be accurately scaled, e.g., statically to conform to hardware capabilities, as I/O instances now \emph{only} perform I/O-related work.
If, as in \sref{sec:arch_static_pool}, hybrid execution is allowed, assumptions on the average I/O to non-I/O work ratio have to be made to estimate the number of I/O instances that would not bottleneck the hardware, with any overshoot in the estimate resulting in reduced CPU efficiency.

Similarly dynamically scaling the amount of active I/O instances based on the current I/O load is now efficiently possible, as with IO threads reducing the amount of active I/O instances does not proportionally reduce the amount of resources available for non-I/O related tasks. 
Deactivating instances can be achieved by adjusting the distribution of requests in the distribution layer to skip the I/O instances in question.
If an I/O instance is consequently starved of requests for long enough, the corresponding kernel thread goes to sleep, and the thread(-pair) of the I/O instance should be implemented to then also go to sleep.
Alternatively the CPU time of the user space threads of an I/O instance could also be provided by the users via a function call, rather then a dedicated thread.
In this case, rather than going to sleep, the thread or thread-pair could simply return.
This approach distributes requests to fewer I/O instances in times of low load and releases the high upkeep cost resources of polling threads of currently unused I/O instances.

%% file: inc/content/summary.tex
\section{Summary}
\label{sec:summary}

While contemporary async APIs, such as \IOURING{}, can significantly improve performance, their adoption both in commercial and academic systems has been limited, likely due to the complexities inherent in achieving adequate performance when using them, which have been laid out in this paper.
We discussed the most notable architectures that may be used for high performance async I/O.
The shared nothing approach, which, although most efficient, is limited in applicability, the static I/O thread pool, which struggles with scaling and CPU efficiency, if otherwise beneficial features such as SQ polling are used and the dynamic I/O thread pool that attempts to address this issue by actively managing the kernel threads introduced by SQ polling.
We hope that our analysis of currently employed asynchronous I/O architectures will spark further discussion and lead to a better adoption and more convenient integration of asynchronous I/O APIs into I/O-intensive applications, such as database systems and I/O libraries.

%% file: main.bbl
\begin{thebibliography}{10}

\bibitem{winIoRings}
ioringapi - win32 apps.

\bibitem{nvmeSpecs}
Specifications - {NVM} express.

\bibitem{amdEpycGen4}
{\sc AMD}.
\newblock 4th gen amd epyc processor architecture.
\newblock White paper, AMD, 5 2024.

\bibitem{iouringIntro}
{\sc Axboe, J.}
\newblock {Efficient IO with io\_uring}, 10 2019.
\newblock [Online; accessed 20-October-2023].

\bibitem{libuv}
{\sc Corretgé, S.~I.}
\newblock libuv, 2024.

\bibitem{IoAPICmp2}
{\sc Didona, D., Pfefferle, J., Ioannou, N., Metzler, B., and Trivedi, A.}
\newblock {Understanding modern storage {APIs}: a systematic study of libaio, {SPDK}, and io\_uring}.
\newblock In {\em Proceedings of the 15th {ACM} {International} {Conference} on {Systems} and {Storage}\/} (New York, NY, USA, June 2022), {SYSTOR} '22, pp.~120--127.

\bibitem{Zig}
{\sc et~al., A.~K.}
\newblock Zig, 2024.

\bibitem{tokioURing}
{\sc et~al., C.~L.}
\newblock tokio-uring, 2024.

\bibitem{tigerbeetle}
{\sc Greef, J.~D.}
\newblock Tigerbeetle, 2024.

\bibitem{syscallsBadMicrokernel2}
{\sc Gu, J., Wu, X., Li, W., Liu, N., Mi, Z., Xia, Y., and Chen, H.}
\newblock Harmonizing performance and isolation in microkernels with efficient intra-kernel isolation and communication.
\newblock In {\em USENIX ATC'20\/} (2020).

\bibitem{Haas23}
{\sc Haas, G., and Leis, V.}
\newblock {What Modern NVMe Storage Can Do, and How to Exploit It: High-Performance I/O for High-Performance Storage Engines}.
\newblock {\em Proc. VLDB Endow. 16}, 9 (2023), 2090–2102.

\bibitem{corobase}
{\sc He, Y., Lu, J., and Wang, T.}
\newblock {CoroBase}: coroutine-oriented main-memory database engine.

\bibitem{kioxiaCd8p}
{\sc Kioxia}.
\newblock Cd8p-r product brief.
\newblock Product brief, Kioxia, 2024.

\bibitem{seastar}
{\sc Kivity, A.}
\newblock Seastar, 2024.

\bibitem{manualAsync}
{\sc Kocberber, O., Falsafi, B., and Grot, B.}
\newblock Asynchronous memory access chaining.
\newblock {\em Proc. VLDB Endow. 9}, 4 (12 2015).

\bibitem{Leis18}
{\sc Leis, V., Haubenschild, M., Kemper, A., and Neumann, T.}
\newblock {LeanStore: In-Memory Data Management beyond Main Memory}.
\newblock In {\em ICDE'18\/} (2018).

\bibitem{xNvme}
{\sc Lund, S. A.~F., Bonnet, P., Jensen, K. B.~A., and Gonzalez, J.}
\newblock I/o interface independence with xnvme.
\newblock In {\em SYSTOR '22\/} (2022).

\bibitem{syscallsBadMicrokernel}
{\sc Mi, Z., Li, D., Yang, Z., Wang, X., and Chen, H.}
\newblock {SkyBridge: Fast and Secure Inter-Process Communication for Microkernels}.
\newblock In {\em EuroSys '19\/} (2019).

\bibitem{phisonGen5Controller}
{\sc Phison}.
\newblock Product brochure ps5026-e26.
\newblock Product brief, Phison, 2024.

\bibitem{IOApICmp}
{\sc Ren, Z., and Trivedi, A.}
\newblock {Performance Characterization of Modern Storage Stacks: POSIX I/O, Libaio, SPDK, and Io\_uring}.
\newblock In {\em CHEOPS '23\/} (2023).

\bibitem{samsungPm1743}
{\sc Semiconductor, S.}
\newblock Pm1743 ssd whitepaper.
\newblock White paper, Samsung Semiconductor, 2021.

\bibitem{syscallsBadSyscallAlternative}
{\sc Soares, L., and Stumm, M.}
\newblock {FlexSC}: Flexible system call scheduling with {Exception-Less} system calls.
\newblock In {\em OSDI'10\/} (2010).

\bibitem{asyncIoCoroutines}
{\sc von Merzljak, L., Fent, P., Neumann, T., and Giceva, J.}
\newblock What are you waiting for? use coroutines for asynchronous {I/O} to hide {I/O} latencies and maximize the read bandwidth!
\newblock In {\em ADMS'22\/} (2022).

\bibitem{Yang17}
{\sc Yang, Z., Harris, J.~R., Walker, B., Verkamp, D., Liu, C., Chang, C., Cao, G., Stern, J., Verma, V., and Paul, L.~E.}
\newblock {{SPDK:} {A} Development Kit to Build High Performance Storage Applications}.
\newblock In {\em CloudCom'17\/} (2017).

\bibitem{Zhou23}
{\sc Zhou, Z., Bi, Y., Wan, J., Zhou, Y., and Li, Z.}
\newblock {Userspace Bypass: Accelerating Syscall-intensive Applications}.
\newblock In {\em OSDI'23\/} (2023).

\end{thebibliography}
